\documentclass[journal=jacsat,manuscript=article]{achemso}
\usepackage{indentfirst}
\usepackage{amsmath}
\DeclareUnicodeCharacter{2212}{-}
\let\oldAA\AA
\renewcommand{\AA}{\text{\normalfont\oldAA}}

\usepackage{chemformula} 
\usepackage[T1]{fontenc} 

\newcommand{\AffISSP}{\affiliation{Institute for Solid State Physics (ISSP), The University of Tokyo, Kashiwa, Chiba 277-8581, Japan}}

\newcommand{\AffIbaraki}
    {\affiliation{Faculty of Pure and Applied Sciences, University of Tsukuba, Tsukuba, Ibaraki 305-8571, Japan}}
\newcommand{\AffAIST}{\affiliation{CD-FMat, The National Institute of Advanced Industrial Science and Technology (AIST), Tsukuba, Ibaraki 305-8560, Japan}}
\newcommand{\AffKyu}{\affiliation{Faculty of Engineering, Kyushu University, Fukuoka, Fukuoka 819-0395, Japan}}
\newcommand{\AffRIKEN}{\affiliation{RIKEN Center for Emergent Matter Science, Wako, Saitama 351-0198, Japan}}

\author{Xiaoni Zhang}\AffISSP
\author{Yuki Tsujikawa}\AffISSP
\author{Ikuma Tateishi}\AffRIKEN
\author{Masahito Niibe}\AffISSP
\author{Tetsuya Wada}\AffISSP
\author{Masafumi Horio}\AffISSP

\author{Miwa Hikichi}\AffIbaraki
\author{Yasunobu~Ando}\AffAIST

\author{Kunio~Yubuta}\AffKyu
\author{Takahiro~Kondo}\AffIbaraki
\author{Iwao~Matsuda}\AffISSP
\email{imatsuda@issp.u-tokyo.ac.jp}
\title[An \textsf{achemso} demo]
  {Electronic topological transition of 2D boron by the ion exchange reaction}


\begin{document}


\newpage
\begin{abstract}
We systematically investigated electronic evolutions of non-symmorphic borophene with chemical environments that were realized by the ion exchange method. Electronic structures can be characterized by the topological ${Z}_2$ invariant. Spectroscopic experiments and DFT calculations unveiled that a sheet of hydrogenated borophene (borophane) is the Dirac nodal loop semimetal (${Z}_2$=-1), while a layered crystal of $\rm YCrB_{4}$ is an insulator (${Z}_2$= 1). The results demonstrate the electronic topological transition by replacement of the counter atoms on the non-symmorphic borophene layer.
\end{abstract}

\newpage
\section{Introduction}

  Discoveries of two-dimensional (2D) layers, such as graphene, have brought into focus on materials where quasiparticles are described in terms of the same Dirac equation that governs behaviors of relativistic particles. Such 2D Dirac materials have been reported to show exotic optical and electronic properties  \cite{MatsudaTextbook,Lee2008,Balandin2011, Ni2007, Zhang2005, Novoselov2005, Nair2008}. In recent times, novel monolayer materials have been pursued and atomic sheets of boron, so-called borophene, have been found as one of the family of 2D Dirac materials\cite{Feng2018, Feng2017, Feng2016, Matsuda2021}. However, the borophene layers were synthesized only on crystal surfaces in ultrahigh vacuum\cite{Mannix2015,Feng2018, Feng2017, Feng2016}. Nowadays, chemical terminations, such as hydrogenations, have been examined to prepare 2D boron layers that can survive under ambient condition. A method of the ion exchange has been successful to extract sheets of hydrogen boride (HB, borophane or hydrogenated borophene) from crystals of metal borides that are composed of boron layers and metal ions\cite{Matsuda2021, Nishino2017, Tateishi2019, Niibe2021, Tateishi2022}. Borophane are formed by processes of liquid exfoliation of layers in the mother crystal, deintercalation of metallic cation, and hydrogenation. Since there are varieties of 2D polymorphs of boron in metal borides, this chemical approach has an advantage to prepare borophane with various atomic structures. 

\begin{figure}[t]
\setlength{\belowcaptionskip}{-0.5cm}  
    \centering
   \includegraphics[width=.85\linewidth,keepaspectratio]{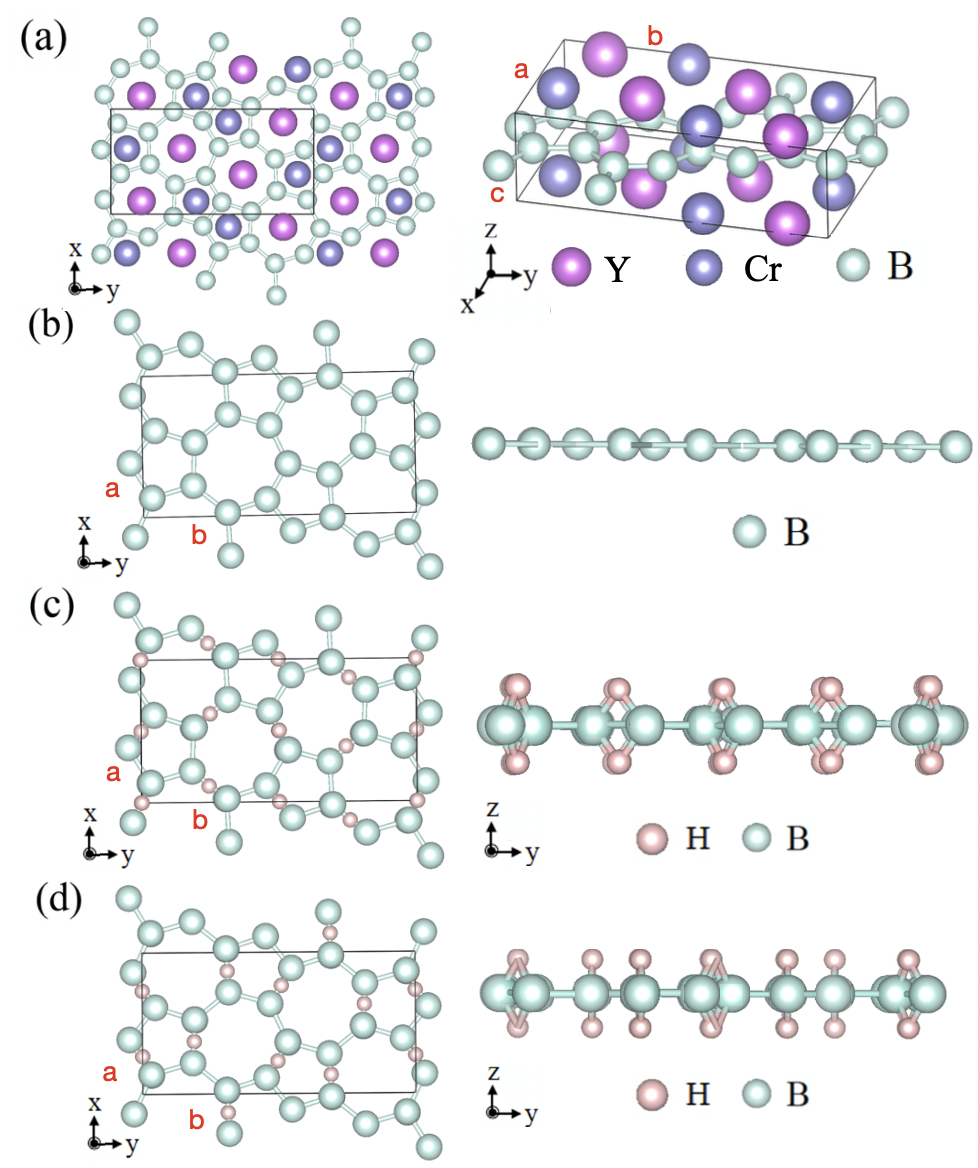}
     \caption{\label{fig1} (color online) Top (left) and side (right) views of atomic structures of (a) $\rm YCrB_{4}$\cite{YCrB4}, (b) 5,7-membered rings borophene (5,7-borophene), (c) 5,7-$\rm \alpha_{1}$-borophane, and (d) 5,7-$\rm \alpha_{2}$-borophane. In each figure, a unit cell is drawn in solid lines with labels of the lattice constants, while atoms are shown as spheres with different colors\cite{VESTA}. }
\end{figure}

  Among metallic borides, a crystal of the $\rm RMB_{4}$-type (R:rare-earth metal, M:metal) has a layered structure with a 2D boron network of the five- and seven-membered rings (5-7-rings), as illustrated in Figs.1(a) and (b). The crystal structures belong to the non-symmorphic symmetry group, the layer group No. 44 {$pbam$} or the 3D space group No. 55 {$Pbam$} , that has been paid attentions as a platform for material design due to generation of the Dirac nodal loops or lines (DNLs) \cite{Guan2017, Dresselhaus2018, Wieder2016, Wieder2018}. The hydrogenated borophene or borophane with 5-7-rings also belongs to the {$Pbam$} group and, as drawn in Figs.1(c) and (d), there are two types of the atomic structures, $\rm \alpha_{1}$ borophane and $\rm \alpha_{2}$ borophane, depending on arrangements of hydrogen atoms\cite{Cuong2020}. Electronic structure of the borophane layer was predicted to be Dirac nodal semimetal by the topological classification based on the ${Z}_2$ invariant and it was confirmed by the DFT band calculation\cite{Cuong2020}. HB sheets, prepared from $\rm TmAlB_{4}$ crystals, were experimentally found to have a gapless electronic structure and those from $\rm YCrB_{4}$, were confirmed to have no metal atom of the mother material after the ion-exchange\cite{Cuong2020}. These physical and chemical characters indicate that the material system is suitable to comprehensively unveil electronic evolutions of the non-symmorphic 2D boron under different chemical environments. 
  
  In the present research, we synthesized borophane layers from crystals of $\rm YCrB_{4}$, prepared from simple substances of Y, Cr, and B. Changes of chemical states of boron were traced by X-ray photoelectron spectroscopy and infrared absorption spectroscopy. Evolutions of electronic states were investigated by DFT calculation and X-ray absorption and emission spectroscopies that probe both occupied and unoccupied states of the matters. Topological classifications by the ${Z}_2$ invariant were carried out for the materials. The detailed procedure is described in the supplementary information. The mother material, a $\rm YCrB_{4}$ crystal, is a trivial insulator (${Z}_2$=1), while the borophane layer is a non-trivial or Dirac nodal semimetal (${Z}_2$=-1). Ion-exchange or replacement of the counter atoms on the non-symmorphic borophene layer was found to exhibit an electronic topological transition.

\begin{figure}[t]
\setlength{\belowcaptionskip}{-0.5cm}  
    \centering
     \includegraphics[width=.5\linewidth,keepaspectratio]{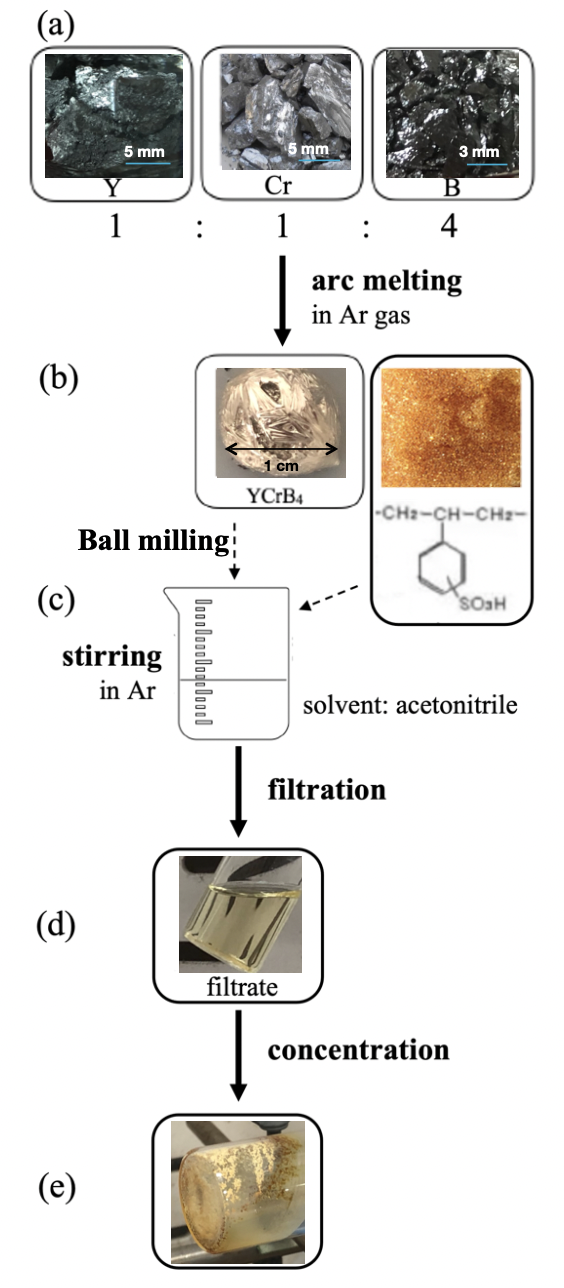}
     \caption{\label{fig2} (color online) Process of synthesizing sheets of borophane. (a) Arc melting to grow $\rm YCrB_{4}$ crystals, (b) ball milling of the crystal and mixing with ion resin, (c) ion exchange reaction, (d) filtrate after filtration, (e) dried sample after concentration.}
\end{figure}

\section{Experiment and calculation} 
\subsection{Synthesis procedures}
Sheets of hydrogen boride or borophane were prepared by the ion-exchange reaction of metal borides with protons that have been successfully made from $\rm MgB_{2}$ crystals\cite{Nishino2017}. Recently, the method was also found to to be applicable for $\rm YCrB_{4}$ crystals\cite{ Niibe2021}. A whole process is described in Fig.2. The mother material, $\rm YCrB_{4}$, was synthesized by the melting method\cite{YCrB4}. Metals of three elements, Y, Cr, and B, were placed on a Cu hearth with atomic ratio of Y:Cr:B=1:1:4, as Fig.2(a),followed by arc melting under the Ar atmosphere. We used yttrium ingots (size:7-10 mm, purity: 99.9 \%),  chromium shots (10-20 mm, 99.999\%), and boron crystals (1-3mm, 99.5\%). The samples were turned over and were remelted three times to ensure homogeneity. A button shaped $\rm YCrB_{4}$ crystal, Fig.2(b), was examined by X-ray diffraction and its crystallinity were confirmed. The experimental lattice parameters, which were almost the same as the previous result \cite{YCrB4}, were used for theoretical calculation.  By ball milling, the sample was divided into pieces and mixed with cation ion-exchange resin beads (0.5-1 mm) in solvent of acetonitrile. Then, in environment of inert gas, Ar, the solution was stirred at room temperature for 1-3 weeks(Fig.2(c)). Hydrogenation of boron sheets in crystals of rare-earth aluminum/chromium boride was conducted by the proton ion-exchange reaction, associated with liquid exfoliation\cite{Nishino2017,Ma2010}. For example, in the case of the yttrium chromium boride ($\rm YCrB_{4}$), the reaction can be described as $n$$\rm YCrB_{4}$ + 4$n$H$\rm ^{+}$ $\rightarrow$ $n$Y$\rm ^{(4-x)+}$ + $nCr$$\rm ^{x+}$ + 4$n$HB, where HB represents hydrogen boride or borophane sheets with a stoichiometric ratio of B : H = 1 : 1. It is worth mentioning that it reaches to higher productive rate for the longer reaction time and smaller $\rm YCrB_{4}$ crystal size. It likely indicates that the ion exchange proceeds at the liquid/solid interface of the crystal. In the ion-exchange method, the borophane sheets are distributed in solvent and can be obtained by filtration. The filtrate and the sample powder, obtained by the drying concentration, are colored in yellow, as shown in Fig.2 (d) and (e), respectively.

\subsection{Measurement conditions}
Analyses of elements and chemical states of samples were examined by X-ray photoelectron spectroscopy (XPS) at the soft X-ray beamline (BL) BL07LSU in the synchrotron radiation (SR) facility, SPring-8 \cite{Yamamoto}. The 5,7-borophane sample (filtrate, Fig. 2(d)), was mounted on a Mo sample holder, then transferred into chamber for XPS measurement. The electronic structures were investigated by photoemission spectroscopy (PES) at the beamline and by a combination of x-ray absorption and emission spectroscopy (XAS/XES) spectroscopies at BL-09A in the NewSUBARU SR facility \cite{Niibe2004, Niibe2016}. It is of note that unoccupied states are probed by XAS, while occupied states by PES and XES. In the XPS and PES measurements, the incident photon beam was linearly polarized at $>99.6\%$. The PES spectra were recorded by an electron spectrometer. The measurement of XAS/XES was held at the B K-shell absorption edge (B K-edge), which allows us to investigate electronic structure selectively at boron sites in a sample. All the measurement were conducted with linearly polarized X-ray and at room temperature. Signal detections of XAS were made by the total electron yield (TEY) and total fluorescence yield (TFY). The excitation energy was set at $h\nu$=210 eV for XES measurements. 

Characterizations of vibrational states were examined by Fourier transform infrared spectroscopy (FTIR) with a commercial system of ALPHA II (BRUKER). The FTIR spectra were recorded by the attenuated total reflectance (ATR) method using a prism holder. FTIR measurements were made at room temperature under argon gas atmosphere.  

\subsection{Theoretical calculation}
Electronic structure of the atomic sheets and the metal boride crystals were calculated in a framework of density functional theory (DFT) by the Quantum ESPRESSO code \cite{qe}. Spin-orbit coupling is neglected and for the exchange-correlation term, the generalized gradient approximation (GGA) with nonrelativistic Perdew-Burke-Ernzerhof parametrization \cite{PBE} is used. Valence wave functions were expanded using a planewave basis and the cutoff energies were set at 60 and 400 Ry for wave functions and charge density, respectively. For PDOS calculation, the gaussian broadening width is set as 0.01 Ry.  The ${k}$-point gild on the Brillouin zone is taken as $16 \times 8 \times 1$ for 2D cases (borophene and borophane) and $8 \times 4 \times 16$ for 3D cases (metal borides) in the band calculations, and $36 \times 18 \times 1$ for 2D cases and $12 \times 6 \times 24$ for 3D cases in the DOS calculations, respectively. Parameters of $\rm YCrB_{4}$ unit cells were taken as a = 5.87519$\rm \AA$, b = 11.29208$\rm \AA$, c = 3.52511$\rm \AA$, which were optimized from experimental values. Lattice parameters of borophene and borophane were taken from the reference\cite{Cuong2020}: a= 5.87519$\rm \AA$, b= 11.29206$\rm \AA$ for the 5,7-borophene, a= 5.87524$\rm \AA$, b=11.29211$\rm \AA$ for the 5,7-$\rm \alpha_{1}$ and $\rm \alpha_{2}$ borophane.

\section{Results and discussion}

\subsection{Chemical characterization}

\begin{figure}[t]
    \centering
     \includegraphics[width=.8\linewidth,keepaspectratio]{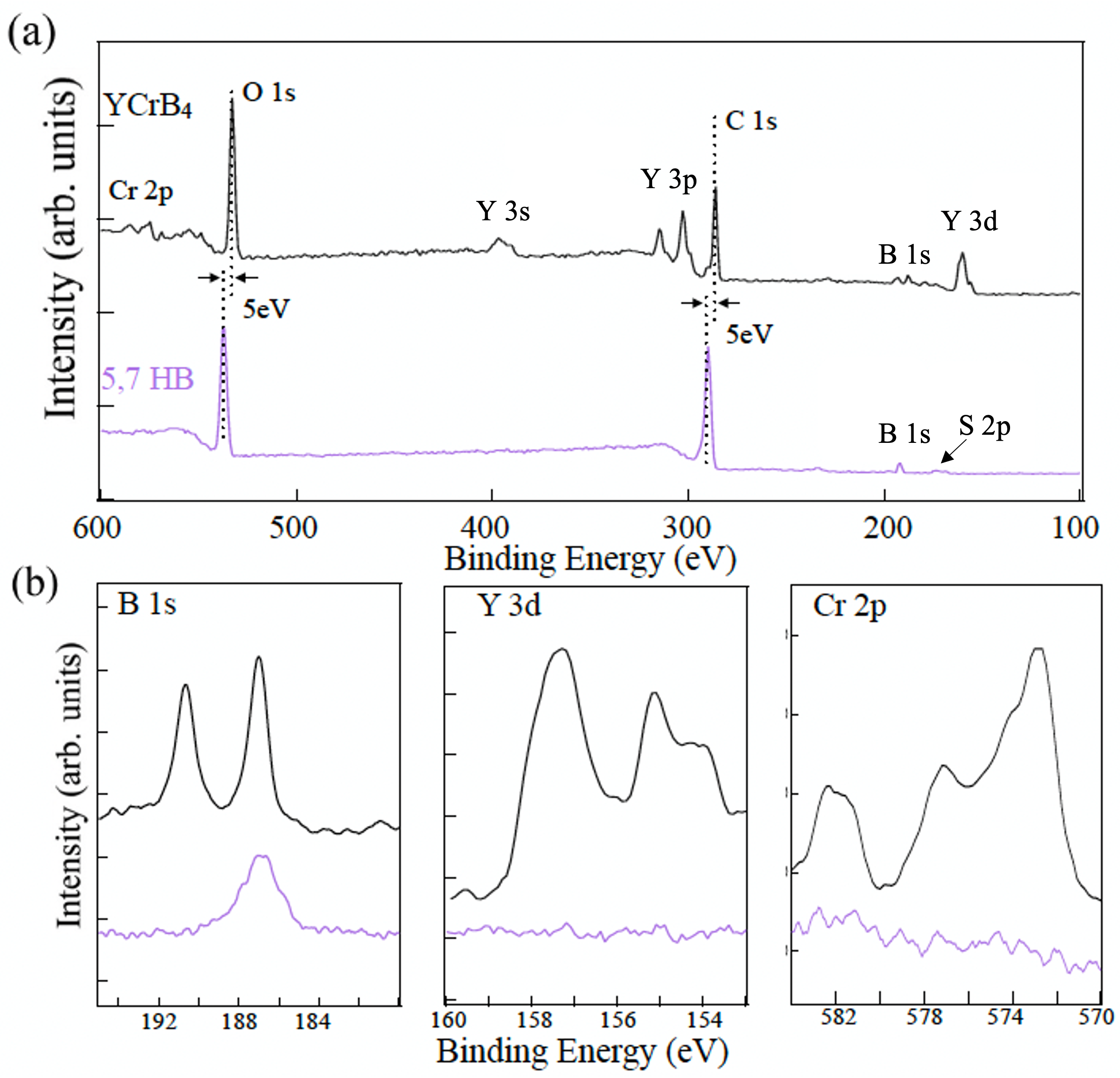}
     \caption{\label{fig3} (color online) (a) A change of the XPS spectra of $\rm YCrB_{4}$ (colored in black) and of 5,7-borophane (purple) by the ion-exchange reaction.
      (b) Comparisons of the core level spectra, B 1s, Y 3d, and Cr 2p, of $\rm YCrB_{4}$ (black) and of 5,7-borophane (purple). }
\end{figure}

Figure 3 (a) compares chemical states and compositions of a sample as XPS spectra before and after the synthesis procedure (Fig.2). A spectrum of the $\rm YCrB_{4}$ crystal (Fig.2 (b)), shows apparent core-level peaks of the composing elements, Y, Cr, and B. In addition, one can identify notable peaks of the carbon and oxygen atoms of impurities at the sample surface that were contaminated by air molecules during mount and transfer. Since these peak positions (C 1s and O 1s) have been defined precisely, the values are used as the energy reference. In Fig.3 (a), an XPS spectrum of the 5,7-borophane sample (Fig.2 (e)), shows absence of all the Y and Cr peaks, indicating that the ion-exchange process, Fig.2 (c), was held successfully. Except the boron atoms, one can find notable peaks of the carbon and oxygen atoms. In addition, core-level peaks of sulfur are observed. These spectral features mean that the borophane sample contains impurities, including pieces of the ion resin. Since a resin is insulating, the sample macroscopically lacks conductivity and induces the charge-up effect in photoemission experiments. As a result, the overall XPS spectra of the sample shifts to the higher binding energy, as shown in Fig.3(a). Accurate binding energy of the individual core-level is evaluated after compensation of the 5 eV shift by the energy position of C 1s and O 1s. Figure 3(b) is a collection of core-level spectra of the $\rm YCrB_{4}$ crystal and the 5,7-borophane sheet, taken at the energy region of B 1s, Y 3d, and Cr 2p. Spectral comparisons confirm complete removal of the Y and Cr atoms by the ion exchange reaction. 

In B 1s spectra, two peaks of the $\rm YCrB_{4}$ sample are found at binding energy of 187.1 eV and 191.2 eV that are assigned to borophene layers in the crystals and impurity boron oxides, respectively\cite{Niibe2021,Moddeman2005}. On the other hand, one peak is detected for the borophane sample, indicating absence of the $\rm BO_{x}$ specie. Recalling XPS data of various boron materials \cite{Il2005, Matsuda2021,Moddeman2005}, energy position of the boron layer implies that the boron atom is negatively charged in both $\rm YCrB_{4}$ and 5,7-borophane. This is consistent to an ionic character of a metal boride that electrons are transferred from metal atoms, i.e. Y or Cr to B atoms. Since the boron atom in the borophane layer shares the similar binding energy, it is natural to conclude that it is also negatively charged and the material behaves as hydrogen boride (HB).

\begin{figure}[t]
\setlength{\belowcaptionskip}{-0.2cm}  
    \centering
     \includegraphics[width=.8\linewidth,keepaspectratio]{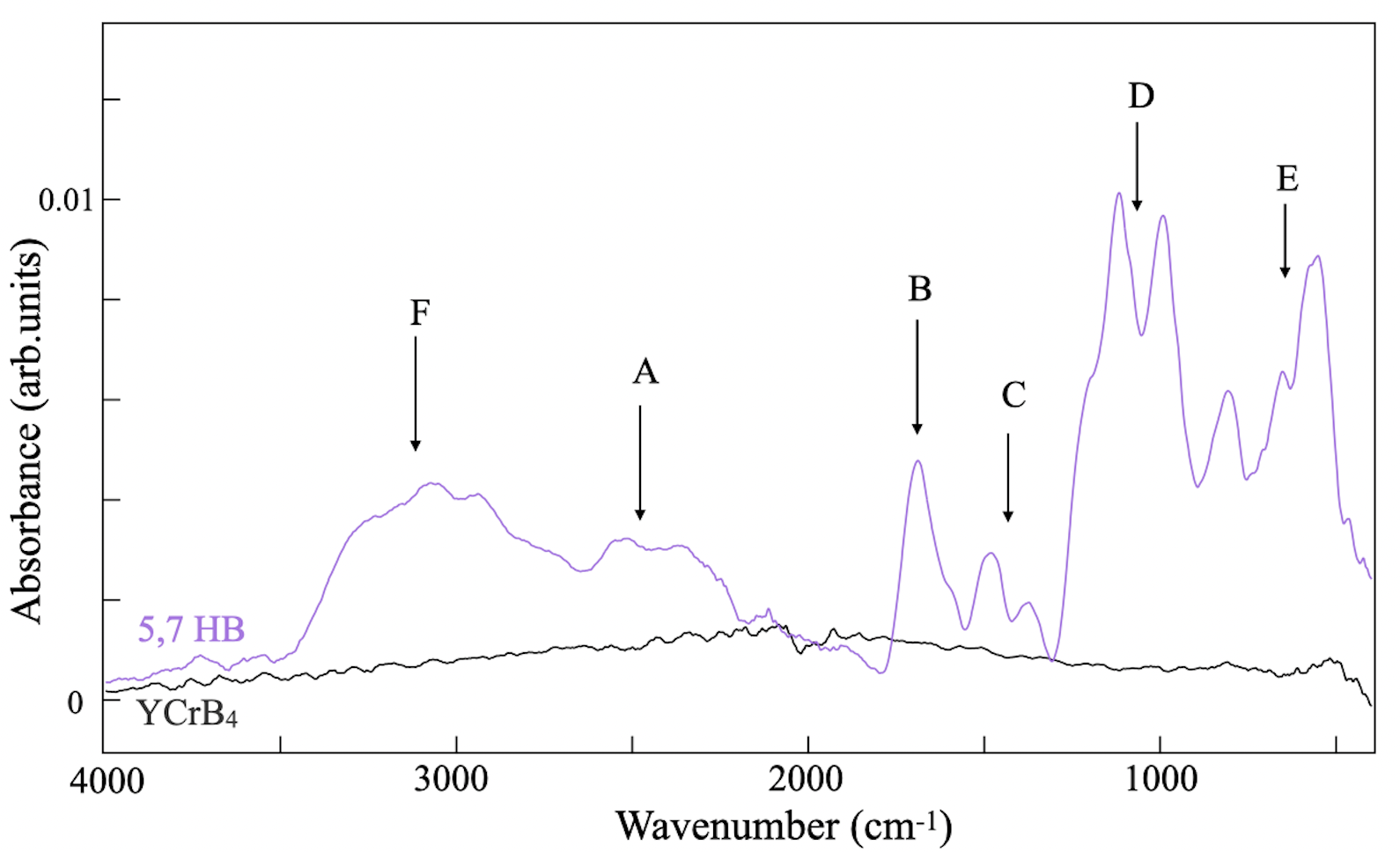}
     \caption{\label{fig4} (color online) FT-IR spectra of $\rm YCrB_{4}$ (black) and of 5,7-borophane (purple).}
\end{figure}

\begin{table}[t][]
 \caption{\label{T1} Assignment of FT-IR absorption .}
\begin{tabular}{|l|l|l|l|}
\hline
Label & \begin{tabular}[c]{@{}l@{}}wavenumber\\ $\rm cm^{-1}$ \end{tabular} & \begin{tabular}[c]{@{}l@{}}reference\\ $\rm cm^{-1}$\end{tabular} & \begin{tabular}[c]{@{}l@{}}assignment\\ vibration mode\end{tabular} \\ \hline
A  &  2450 & 2500& B-H terminal stretching\cite{Tominaka}\\\hline
B  & 1691 & 1619 & B-H-B linkage \cite{Nishino2017}\\\hline
C  & 1490 & 1400 & B-H-B bridging\cite{Tominaka} \\\hline
D  & 1120 & 1000 & B-H stretching\cite{Kawamura2020, Kawamura2019} \\\hline
E  & 661 & 700 & B-H stretching\cite{Nishino2017}\\\hline
F  & 3179 & 3200 & O-H stretching\cite{Kawamura2019} \\\hline
\end{tabular}
\end{table}

To examine hydrogenation of the boron layer, Figure 4 shows FTIR absorption spectra of the samples taken at a range of 450-4000 $cm^{-1}$ that covers absorption by vibrations at various B-H bonds. One can find characteristic absorption features, labeled A-E, for the 5,7-borophane sample that correspond to B-H or B-H-B vibrational modes. The assignments can be made by previous FTIR research of the honeycomb borophane \cite{Nishino2017, Kawamura2020, Kawamura2019, Tominaka}, as summarized in Table I. The results evidence hydrogenation of borophene and formation of borophane (HB). The absorption, labeled F, is due to O-H stretching vibration mode of water molecules adsorbed on the HB sheets \cite{Kawamura2019, Kawamura2020}.

\subsection{Electronic structure analysis}

\begin{figure}[t]
\setlength{\belowcaptionskip}{-0.2cm}  
    \centering
     \includegraphics[width=.8\linewidth,keepaspectratio]{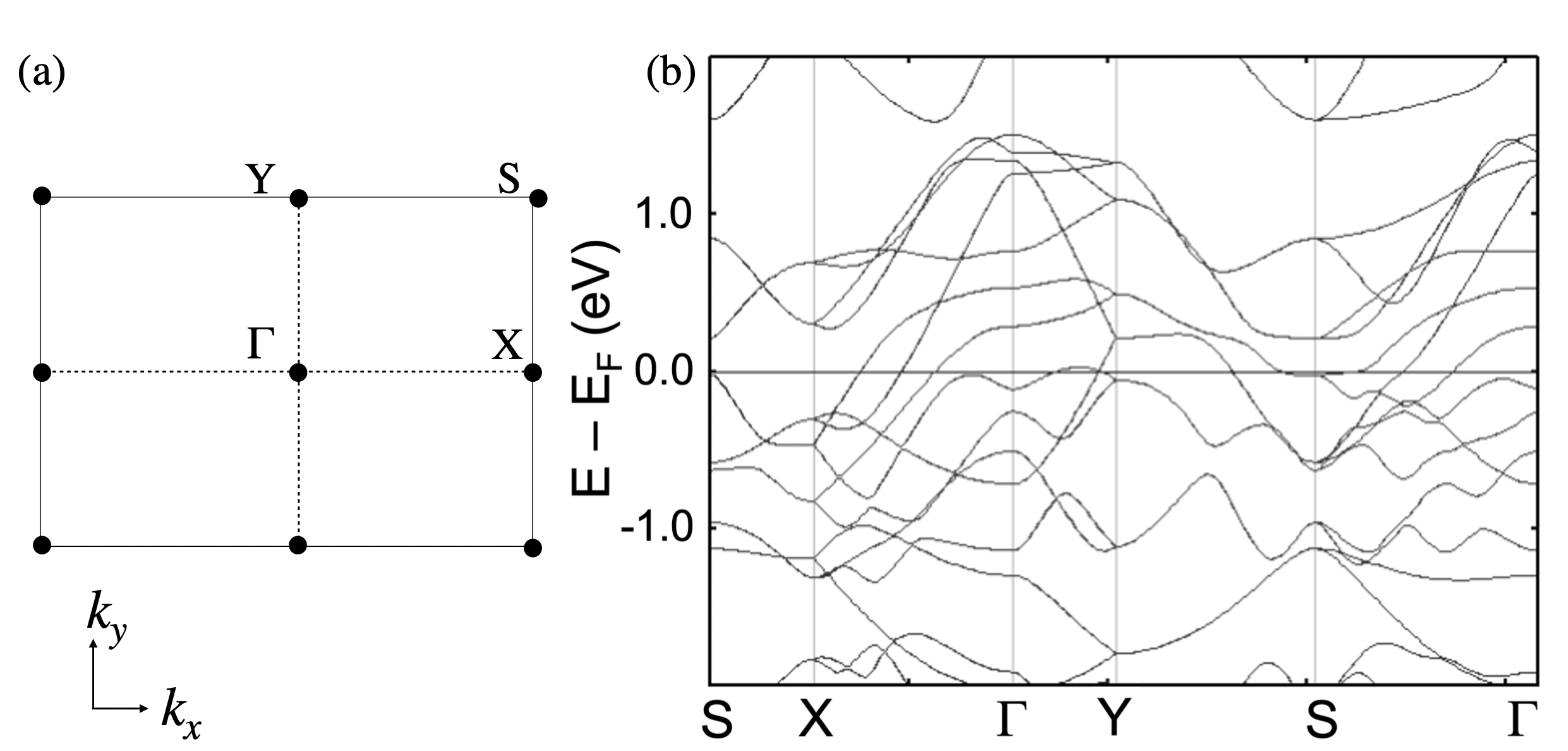}
     \caption{\label{fig5} (color online) Band dispersion curves of a freestanding layer of 5,7-borophene along the k paths in the Brillouin zone. The Fermi level, $E_F$, position is outlined by a black solid line. }
\end{figure}

Procedure of the ion exchange reaction from $\rm YCrB_{4}$ to 5,7-borophane corresponds to variation of chemical environment of 5,7-borophene. Based on results of X-ray spectroscopy and DFT calculation, we discuss the electronic evolutions of the boron layers. Figure 5 shows calculated dispersion curves of bands of a pristine 5,7-borophane. The layer is metallic and it has many bands around the Fermi level ($E_F$).

Partial density-of-states (PDOS), shown in Fig.6(c), consistently reveals that a large weight of both the $\sigma (\sigma^{*})$ and $\pi (\pi^{*})$ orbitals exists at $E_F$. It is of note for a clarity of our discussion, we define $\sigma (\sigma^{*})$ orbitals for the in-plane $p$ orbitals ($\rm p_{x}$, $\rm p_{y}$), while the $\pi (\pi^{*})$ orbitals the out-plane $p$ orbitals ($\rm p_{z}$). When the layers generates a crystal with Y and Cr atoms, the large PDOS of $\sigma (\sigma^{*})$ and $\pi (\pi^{*})$ energetically shift to the occupied state region, Fig.6(b). This is naturally explained in terms of electron transfer from metal atoms to the boron layer, as revealed by the core-level data in Fig.3. Interactions between metal and boron atoms further induce formation of an energy gap and a crystal of $\rm YCrB_{4}$ is semiconducting, as confirmed by XAS/XES data (Fig.6(a)) and by calculation (Fig.6(b)). Amount of the energy gap is evaluated 200 meV from the experiments and 150 meV by the DFT calculation. The difference is reasonable since the GGA generally tends to underestimates the band gap. On the other hand, two types of 5,7-borophane layers, $\rm \alpha_{1}$-type and $\rm \alpha_{2}$-type, have gapless electronic structure at $E_F$. By comparing with PDOS of a pristine 5,7-borophane, one finds that the hydrogenation makes the $\sigma$-type band almost fully occupied and generates large splitting of the $\pi$-type band by the B-H-B bond formation. As a consequence, the electronic states energetically overlap with each other at $E_F$. The spectral feature corresponds to electronic states of Dirac nodal semimetal, as reported previously \cite{Niibe2021} . 

\begin{figure}[t]
\setlength{\belowcaptionskip}{-0.2cm}  
     \includegraphics[width=.75\linewidth,keepaspectratio]{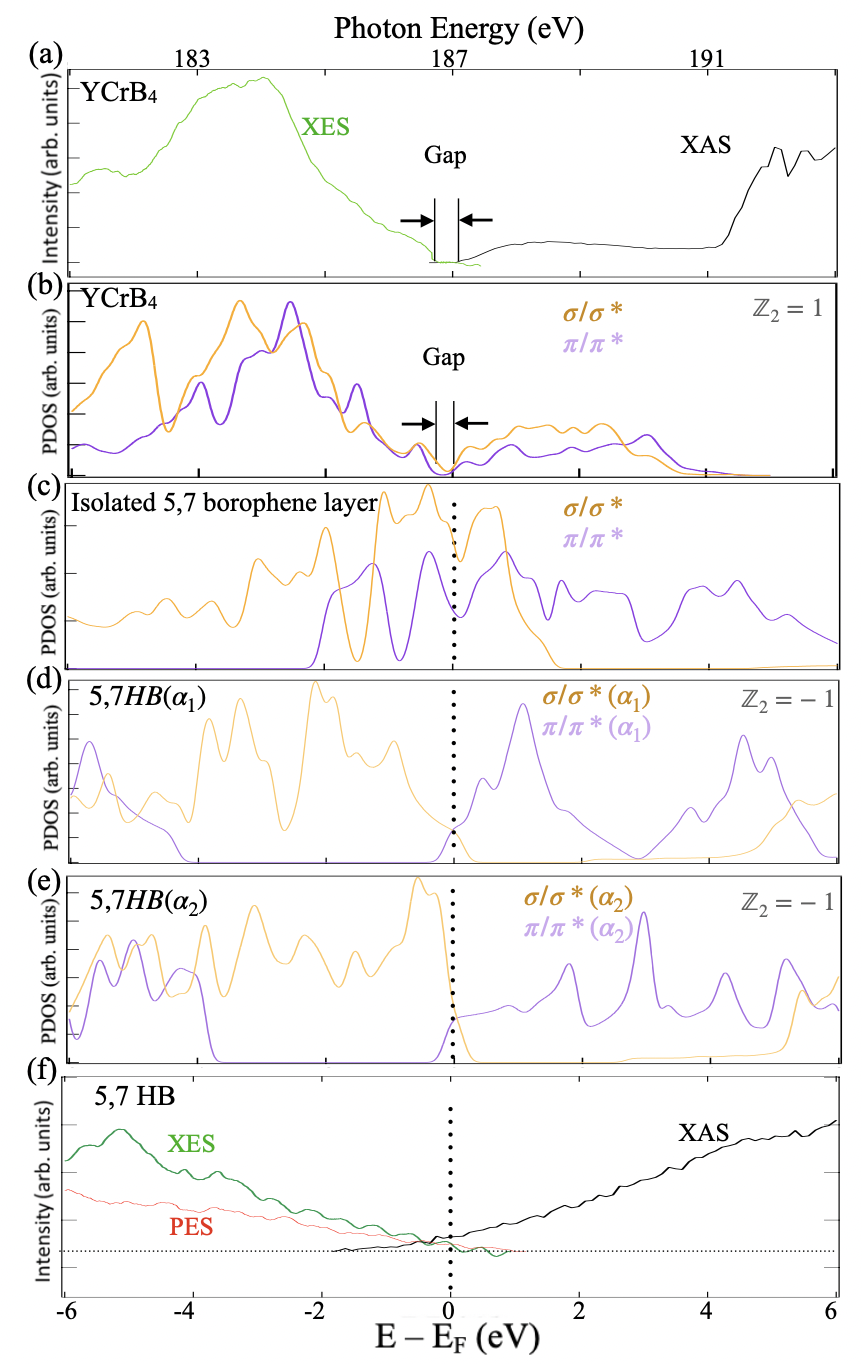}
     \caption{\label{fig6} (color online) Evolutions of occupied and unoccupied states between (a,b) boron layer in $\rm YCrB_{4}$, (c) borophene (B), and (d-f) borophane (HB). Experimental spectra of XAS, XES, and PES are given in (a,f), while calculated PDOS are shown in (b-e). In the figure, the topological ${Z}_2$ invariants of the materials are indicated.  
     }
     \end{figure}

The gapless or the semimetallicity  feature of the 2D material was evidenced by experimental results of XAS, XES, and PES in Fig.6(f). The XAS spectrum was recorded by the TEY method since it is surface-sensitive. The spectral overlap of XAS (unoccupied state) and XES (occupied state) can be identified at 187 eV that corresponds to $E_F$, as expected from the PDOS in Fig.6(d) and (e). It is of note that the spectrum of our borophane sample shared similar features with the $\rm \alpha_{1}$-type structure. In previous report, a 5,7-borophane synthesized from $TmAlB_{4}$, also showed consistency with $\rm \alpha_{1}$-type structure.\cite{Niibe2021}. The PES spectrum was energetically corrected by the 5 eV-shift, as described in Fig.3. The spectral tail extends up to $E_F$ and confirms existence of gapless electronic states of the 5,7-borophane layer.

A series of electronic states of the boron layers unveils contrasting behaviors between borides that depend on partner elements. In metal borides, i.e. $RMB_{4}$ with R=Y and M=Cr, electrons are doped uniformly into electronic bands of the boron layer, while, in HB, doping was held non-uniformly. The distinction is likely due to the different nature of local chemical bonds between a boron atom and the counterpart. The bonding scheme is rather ionic in metal borides but it is rather covalent and also specific to boron materials. The electron doping in the HB sheet is held in the in-plane band, associated with formation of the B-H-B bonds. 

Semimetallicity of the HB sheet is associated with a Dirac nodal loop and it can be described in terms of the topological ${Z}_2$ invariant, given by a product of the band parity at the $\Gamma $ point \cite{Cuong2020} . A layer of the $\rm \alpha_{1}$ or $\rm \alpha_{2}$ borophane has ${Z}_2$ values of ${Z}_2$=-1, and it is topologically non-trivial. While the detailed arguments were carried out for the atomic sheets, the band topology have never been discussed for the borophene sheet in the mother material, $\rm YCrB_{4}$. Thus, it is of interest to characterize the topological nature and to examine the variation by the ion-exchange reaction. As shown in Fig.6 (a) and (b), a $\rm YCrB_{4}$ crystal is semiconducting and it has an energy gap at $\rm E_F$. The band dispersion plot is given in the Supplementary. A value of the ${Z}_2$ invariant of $\rm YCrB_{4}$ is ${Z}_2$=1, and it means that the materials is topologically trivial. The different ${Z}_2$ values indicates that the ion-exchange reaction made the electronic topological transition. It is of note that it is not able to abruptly define the ${Z}_2$ quantity for a freestanding boron layer due to existence of multiple metallic bands at $\rm E_F$. It is intriguing to find that the ${Z}_2$ values are rigorously defined by the counterpart elements that terminate both sides of the layer. 

\section{Conclusion} 
 
In summary, we systematically characterize the topological nature of 2D borophane sheet with a 5, 7-rings structure under different chemical environment. We realized 2D 5,7-rings borophane by ion-exchange method. The formation of 5,7-rings borophane from metallic boride $\rm YCrB_{4}$ was confirmed by XPS and FT-IR spectra. The electronic states of 5,7-borophene layer in metallic boride and hydrogenated sheet were investigated by photoemission, absorption and emission spectroscopy (PES, XAS and XES) at the B K-edges. The experimental results were reproduced by PDOS calculations using the density functional theory (DFT). 

The current work shows that 5,7-borophane is a semimetal with DNL, while $\rm YCrB_{4}$ is a narrow gap semiconductor. The electronic topological transition of 2D boron layer is created by chemical ion-exchange method. Our results suggest a promising way to design and modify novel 2D topologically non-trivial materials.

\section{Supporting Information} 
Brillouin zone and band dispersion of $\rm YCrB_{4}$ (with and without Spin-orbit coupling effect), Brillouin zone and band dispersion of 5,7-rings borophane ($\rm \alpha_{1}$ and $\rm \alpha_{2}$ structure), evaluation of existence of Dirac Nodal loop in the case of $\rm YCrB_{4}$.

\section{Acknowledgments} 
Rieko Ishii is acknowledged for her technical support of the arc melting. This work was supported by JST, CREST Grant Number JPMJCR21O4, Japan, and by JSPS KAKENHI Grant Numbers JP19H04398, JP18H03874, JP19K05643, JP20H05258 and JP21H00015:B01 .



\newpage
{\centering\section* {Supporting Information: Electronic topological transition of 2D boron by the ion exchange reaction}}



\section{S1. Band diagram of a $\rm YCrB_{4}$ crystal}

\begin{figure}[t]
\setlength{\belowcaptionskip}{-0.2cm}  
    \centering
     \includegraphics[width=.85\linewidth,keepaspectratio]{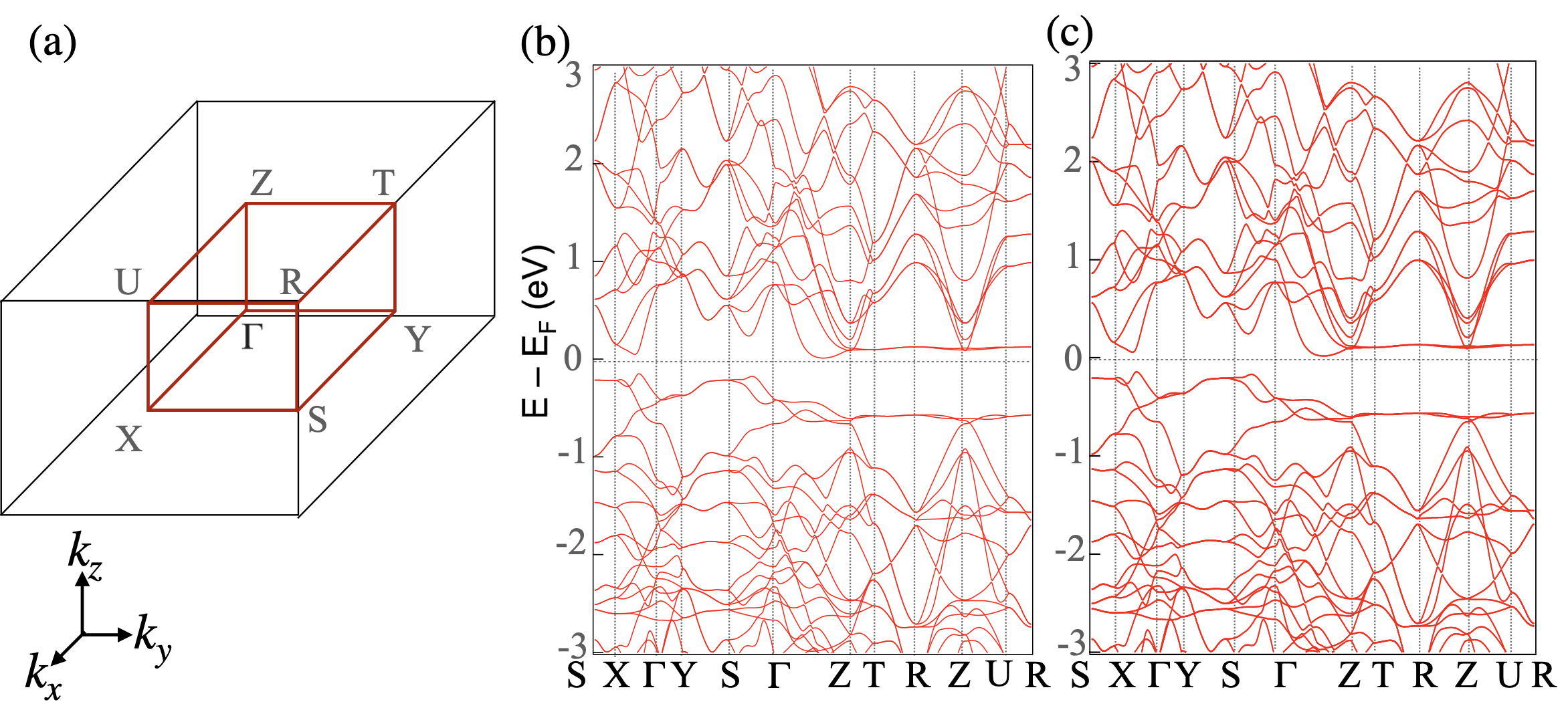}
     \caption{\label{Fig. S1} (color online)   (a) Brillouin zone. (b,c) Band dispersion curves of a $\rm YCrB_{4}$ crystal (b) without spin-orbit coupling (SOC) and (c) with SOC.}
\end{figure}

 A crystal of $\rm YCrB_{4}$ belongs to the non-symmorphic space group $Pbam$, 3D space group No.55. The 3D Brillouin zone of $\rm YCrB_{4}$ is shown in Fig. 7 (a), while the band diagram is given in Fig. 7 (b,c) (b) with Spin-Orbit Coupling (SOC) and (c) without SOC. An indirect energy gap about 150 meV is observed, which corresponds to the PDOS calculation result in Fig.6(b) in the main text. 

\section{S2. Band diagrams of 5,7-borophane layers of the $\rm \alpha_{1}$ and $\rm \alpha_{2}$ types}

\begin{figure}[t]
    \centering
     \includegraphics[width=.8\linewidth,keepaspectratio]{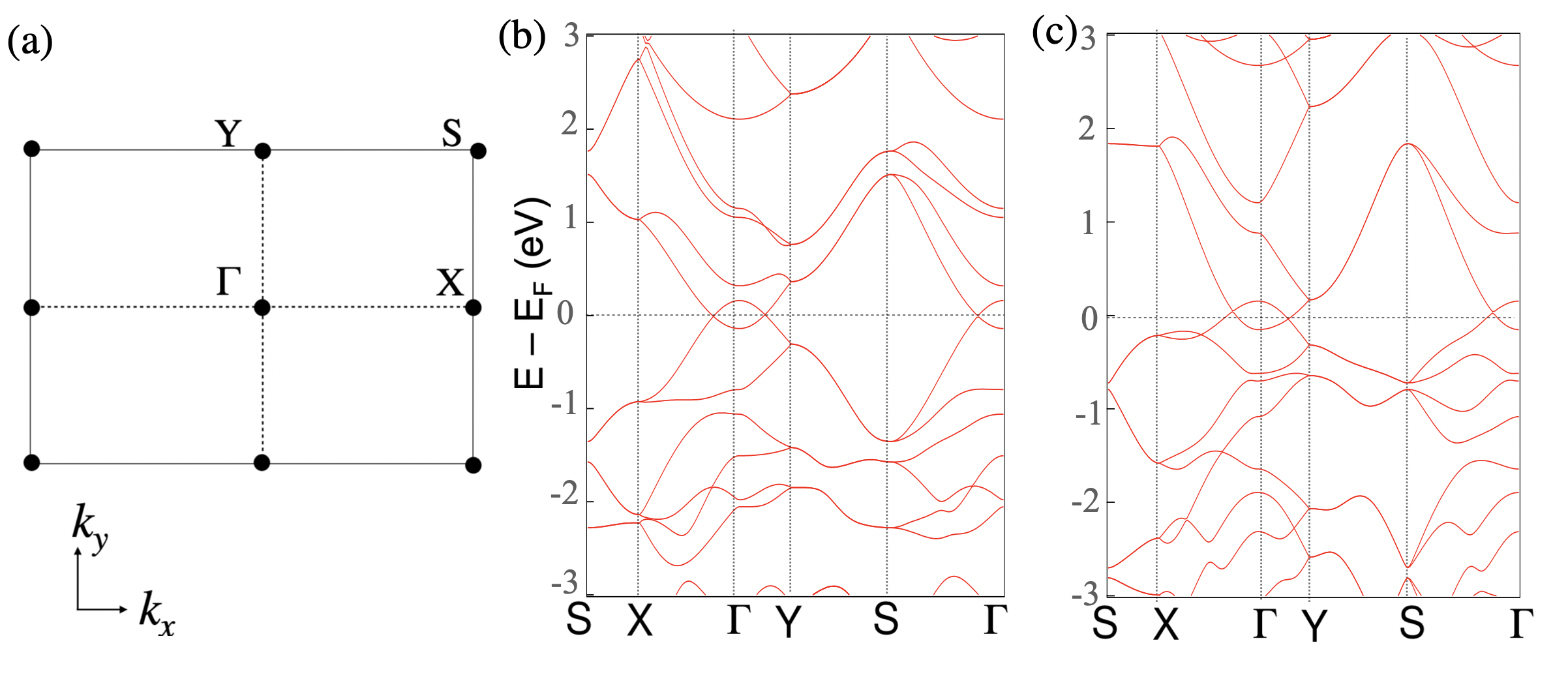}
     \caption{\label{Fig. S2} (color online) (a) Brillouin zone and (b,c) band dispersion curves of 5,7-borophane layers with the $\rm \alpha_{1}$ and $\rm \alpha_{2}$ structures.}
\end{figure}

Layers of 5,7-borophane, the $\rm \alpha_{1}$ and $\rm \alpha_{2}$ structures, belong to the non-symmorphic symmetry group No. 44, $Pbam$. The 2D
Brillouin zone is shown in Fig. 8(a). The electronic structures of two types of borophane, $\rm \alpha_{1}$ and $\rm \alpha_{2}$, are displayed in Fig. 8 (b) and (c), respectively. The Dirac Nodal Loops at the Fermi level correspond to the gapless PDOS in Fig.6(d,e).

\section{S3. Evaluation of existence of DNLs}

In this section we presents the evaluation of DNLs for $\rm YCrB_{4}$ as example. The discussion is followed by two steps:

1. Check compatibility condition of band structure

2. Check symmetry-based indicator

\subsection{S3-1. Band structure compatibility condition}

The compatibility condition depends on the irreducible representations of the occupied energy bands in the high symmetry points (HSPs) (both ends of high-symmetry lines/HSLs). If the number of bands with same eigenvalue is constant, the compatibility condition is satisfied and the system is gapped, as Fig. 9(a), (b) shown.  If the number of bands with same irreps changes on the HSLs, the compatibility condition is violated and the system is gapless as shown in Fig. 9(c). For that sense, compatibility condition can demonstrate the energy band crossings between the HSPs. The previous report essentially tests the compatibility condition while providing a ${Z}_2$ index calculation on 5,7-rings borophane \cite{Cuong2020}.

When the compatibility condition is satisfied, the following situations appears: 

(1). At or between high symmetry points (HSPs), there is no band crossings. 

(2). Along high-symmetry lines (HSLs), band crossings can be gapped without changing the bands order between the occupied bands and unoccupied bands at HSPs. 

Based on above, we discuss the case of $\rm YCrB_{4}$. There is no gapless point along HSLs, and a clear band gap opens at the Fermi level. Therefore, in the case of $\rm YCrB_{4}$, it is clear to show it satisfies the compatibility condition. We also can calculate the topological ${Z}_2$ index by checking Irreps at the $\Gamma$  point, described in S3-3.

\begin{figure}[t]
\setlength{\belowcaptionskip}{-0.2cm}  
    \centering
     \includegraphics[width=.8\linewidth,keepaspectratio]{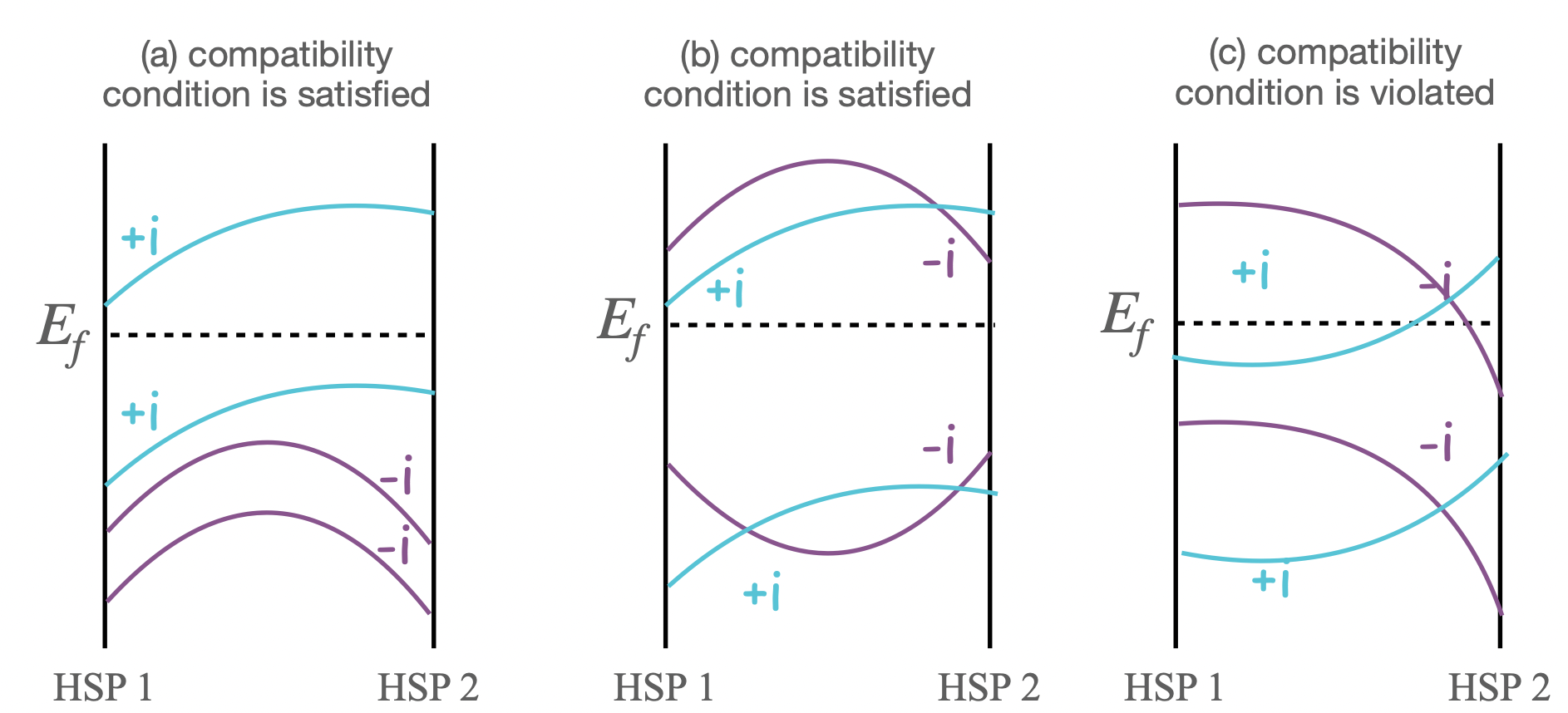}
     \caption{\label{Fig. S3} (color online) Three compatibility conditions along a high-symmetry line. The $\pm i$ is the eigenvalue for the wave function of corresponding band. (a) (b), the number of bands with +i or -i stay constant, the compatibility condition is satisfied and the system is gapped. (c), there is a variation of number of bands with same eigenvalue, the compatibility condition is violated and system is gapless. 
}
\end{figure}

\subsection{S3-2. Symmetry-based indicator}

In an inversion symmetric spineless system, in another word, a system that spin-orbital coupling is negligible, a pair of DNLs can exist at generic points in momentum space even if the compatibility condition is satisfied. These inversion-protected nodal lines do not intersect a HSL, which makes them hard to be found by DFT calculation. In this case, 
We can check the symmetry-based indicator, according to the previous report \cite{PRX}. The symmetry-based indicator can be tested under the situation that compatibility condition is satisfied. 

For the case of $\rm YCrB_{4}$, it satisfies the rule that SOC is neglible (as shown in Fig. 7 (c)), we can check the symmetry-based indicator for $\rm YCrB_{4}$ to detect possible Dirac nodes. In space group No. 55, it is clear to find no non-trivial indicator is defined. Therefore, no inversion-protected DNLs exist in $\rm YCrB_{4}$.

\subsection{S3-3. A brief description of how to calculate the ${Z}_2$ index at $\Gamma $ point}

According to the representation theory of space group, the energy at zone boundary should only be doubly degenerate. With mirror operation for the x-y plane, the number of bands at zone boundary will always be even. 

Generally, in a trivial insulator system, the number of occupied bands at HSPs should stay consistent with the zone boundary, otherwise, a band crossing will be observed around HSP. 

Based on above principles, in the case of $\rm YCrB_{4}$ with $\sigma_{z}$ symmetry operator, if it is a non-metallic system, the number of bands with +1(-1) at $\Gamma $ point should be even, same as the zone boundary. And the indicator of existence of Dirac node is described as topological index ${Z}_2$, with the formula\cite{Cuong2020}: 

\begin{equation}[t]
{ Z}_{2}=\prod_{\rm occupied} \Gamma_i(\sigma_z)= \left \{
\begin{array}{l}
\quad1\quad(\rm even)\qquad\\
-1\quad(\rm odd)\qquad
\end{array}
\right.
\end{equation}

We can obtain the Character table, as shown in Table 1\cite{qe} . 

\begin{table}[][t]
 \caption{\label{Table-S1} Character table of $\rm YCrB_{4}$}
\begin{tabular}{|l|l|l|l|l|l|l|l|l|}
\hline
Label & \begin{tabular}[c]{@{}l@{}}E \end{tabular} & \begin{tabular}[c]{@{}l@{}}$C_{2z}$\end{tabular} & \begin{tabular}[c]{@{}l@{}}$C_{2x}$\end{tabular} & \begin{tabular}[c]{@{}l@{}}$C_{2y}$\end{tabular}& \begin{tabular}[c]{@{}l@{}}I\end{tabular}& \begin{tabular}[c]{@{}l@{}}$\sigma_{z}$\end{tabular}& \begin{tabular}[c]{@{}l@{}}$\sigma_{x}$\end{tabular}& \begin{tabular}[c]{@{}l@{}}$\sigma_{y}$\end{tabular}\\ \hline
$A_{g}$  &  1.00 & 1.00& 1.00& 1.00& 1.00& 1.00& 1.00& 1.00 \\\hline
$B_{1g}$ &  1.00 & 1.00& -1.00& -1.00& 1.00& 1.00& -1.00& -1.00 \\\hline
$B_{2g}$ &  1.00 & -1.00& 1.00& -1.00& 1.00& -1.00& 1.00& -1.00 \\\hline
$B_{3g}$ &  1.00 & -1.00& -1.00& 1.00& 1.00& -1.00& -1.00& 1.00 \\\hline
$A_{u}$ &  1.00 & 1.00& 1.00& 1.00& -1.00& -1.00& -1.00& -1.00 \\\hline
$B_{1u}$ &  1.00 & 1.00& -1.00& -1.00& -1.00& -1.00& 1.00& 1.00 \\\hline
$B_{2u}$ &  1.00 & -1.00& 1.00& -1.00& -1.00& 1.00& -1.00& 1.00 \\\hline
$B_{3u}$ &  1.00 & -1.00& -1.00& 1.00& -1.00& 1.00& 1.00& -1.00 \\\hline
\end{tabular}
\end{table}

From Table 1, the Irreps with $\sigma_{z} = -1$ are $B_{2g}, B_{3g}, A_{u}, B_{1u}$. The total number of bands at $\Gamma $ point, with $\sigma_{z} = -1$ is 18, and the index ${Z}_2$ is:

\begin{equation}[t]
{ Z}_{2}=\prod_{\rm occupied} \Gamma_i(\sigma_z)= (-1)^{18}=1
\end{equation}

Therefore, it is clear to see that $\rm YCrB_{4}$ has no Dirac node at Fermi level. 



\end{document}